\documentclass[showpacs,twocolumn,amsmath,amssymb]{revtex4}
\usepackage{graphicx} 
\usepackage{color}
\usepackage{amssymb,amsfonts,amsmath}
\usepackage{pdfsync}

\begin{document}

\title{Double scaling in the relaxation time in the $\beta$-FPUT model} 

\author{Yuri V. Lvov$^{1}$}
\author{Miguel Onorato$^{2,3}$}
\affiliation{
$^1$ Department of Mathematical Sciences, Rensselaer Polytechnic Institute, Troy, New York 12180, USA;\\
$^2$Dip. di Fisica, Universit\`{a} di Torino, Via P. Giuria, 1 - Torino, 10125, Italy; \\
$^3$ INFN, Sezione di Torino, Via P. Giuria, 1 - Torino, 10125, Italy
}
\begin{abstract}
We consider the original $\beta$-Fermi-Pasta-Ulam-Tsingou
($\beta$-FPUT) system; numerical simulations and theoretical arguments
suggest that, for a finite number of masses, a statistical equilibrium
state is reached independently of the initial energy of the system.
Using ensemble averages over initial conditions characterized by
different Fourier random phases, we numerically estimate the time
scale of equipartition and we find that for very small nonlinearity it
matches the prediction based on exact wave-wave resonant interactions
theory. We derive a simple formula for the nonlinear frequency
broadening and show that when the phenomenon of overlap of frequencies
takes place, a different scaling for the thermalization time scale is
observed. Our result supports the idea that Chirikov overlap criterium
{ identifies} a transition region between two different
relaxation time scaling.
 \end{abstract}
\maketitle In 1923 at the age of 22 E. Fermi published one of his
first papers \cite{fermi1923dimostrazione} in which the goal was to
show that Hamiltonian systems are in general quasi-ergodic.  At that
time, the paper was considered interesting by the scientific
community; however, it appeared later that the hypotheses needed for
the proof are very restrictive
(\cite{fermi1962collected,gallavotti2001meccanica}).  About thirty
years later Fermi, in collaboration with Pasta, Ulam and Tsingou (see
\cite{dauxois2008fermi} for a discussion on the role played by
Tsingou), came back to the problem using a numerical approach.  The
goal was to study a simple mechanical system and verify that a small
nonlinearity would be enough to let the system reach a thermalized
state. Their research was also motivated by the work of Debye who in
1914 conjectured that normal (in accordance to the macroscopic Fourier
law) heat conduction in solids could be obtained only in the presence
of nonlinearity, see \cite{lepri2016thermal} for recent developments.
They simulated a system of harmonic oscillators perturbed by a cubic
($\alpha$-FPUT) or quartic potentials ($\beta$-FPUT). The results they
obtained numerically \cite{fermi1955studies} were very different from
expectations: instead of observing the equipartition of linear energy,
they observed a recurrent phenomena known as the FPUT-recurrence. This
unexplained result triggered a surge of scientific activity and 
lead to the discovery of solitons \cite{zabusky1965} and
integrability in infinite dimensional systems
\cite{gardner1967method}. However, the FPUT system is only close to an
integrable one \cite{benettin2013fermi} and soliton interactions are
not elastic.

At the same time the simulations of the FPUT system were performed,
Kolmogorov enunciated the KAM theorem which loosely speaking
describes how in a perturbed integrable Hamiltonian system the KAM
tori survive if the perturbation is sufficiently small. Chirikov and
Izraielev \cite{izrailev1966statistical} developed a method for
estimating the threshold of initial energy above which the KAM tori
are destroyed. The basic idea is the following: in the presence of
nonlinearity, linear frequencies are perturbed and, if the
perturbation is larger than the frequency spacing (distance between
two adjacent linear frequencies), then the trajectory may oscillate
chaotically between the two frequencies. This idea, known also as the
Chirikov overlap criterium, is very helpful but not rigorous. Indeed,
for example, there exists a counter example: for the Toda lattice (or
other integrable system) a threshold can be derived but the system is
integrable, therefore never chaotic. The idea of Chirikov and Izrailev
has been followed and different numerical studies confirmed the
presence of a threshold above which the FPUT system reaches a fast
thermalized state (see for example
\cite{livi1985equipartition,casetti1997fermi,DeLuca1999} for a study
on $\beta$-FPUT). However, more recently, numerical simulations of the
$\alpha$-FPUT \cite{ponno2011two,onorato2015} have shown that even for
small nonlinearity the system does reach a thermalized state.  The
explanation of this result was given in \cite{onorato2015}
where it has been shown that for the finite dimensional system of
certain size, six-wave resonant interactions are responsible for
equipartition and only after very long time the system reaches a
thermalized state.

In this Letter we perform a detailed study of the $\beta$-FPUT with a
finite number of masses and, as a first result, we show that, as for
the $\alpha$-FPUT, the weak nonlinear regime is dominated by discrete
six-wave resonant interactions which are responsible for 
thermalization.  Such thermalization seems to occurs for any, even
extremely small, levels of nonlinearity.  We then estimate
the time scale it takes to reach equipartition, and we confirm the result 
numerically. Moreover, we construct
numerically the dispersion relation curve and show that equipartition
is observed also in the condition of no-overlap of frequencies.  By
writing the equation of motion in angle-action variables and by using
the Wick decomposition, we find an explicit formula for the broadening
of the frequencies. When such broadening is larger than the spacing
between frequencies, the Chirikov regime is observed. Therefore, the
Chirikov criteria { identifies a threshold for a more effective mechanism
  of thermalization.}  Consequently, there is a double time scaling to
reach equipartition as a function of the nonlienarity parameter. Our
results are fully supported by numerical simulations.

{\it The model-} We consider the Hamiltonian for a chain of $N$
identical particles of mass $m$ of the type:
 \begin{equation}
H=H_2+H_4
\end{equation}
with 
\begin{equation}
\begin{split}
&H_2=\sum_{j=1}^N\left(\frac{1}{2 }p_j^2+\frac{1}{2}(q_j-q_{j+1})^2\right),\\
&H_4=\frac{\beta}{4}\sum_{j=1}^N(q_j-q_{j+1})^4.
\label{H_FPU}
\end{split}
\end{equation}
$q_j(t)$ is the displacement of the particle $j$ from its equilibrium
position and $p_j(t)$ is the associated momentum; $\beta$ is the
nonlinear spring coefficient (without loss of generality, we have set
the masses and the linear  spring constant equal to 1).  

{\it Analytical Results-} Before performing numerical simulations of
the equations associated to the Hamiltonian (\ref{H_FPU}), we first
outline the derivation of some important theoretical predictions:
    i) the nonlinear correction to the linear frequency,
    ii) the broadening of the frequencies in the presence of
    nonlinearity, iii) the time scale of equipartition. Those
    ingredients will help us in interpreting the numerical results.

Assuming periodic boundary conditions and the standard definition of
the Discrete Fourier Transform, we introduce the following normal
variable as
\begin{equation}
a_k=\frac{1}{\sqrt{2 \omega_k}}(\omega_k Q_k+i P_k),\label{NormalMode}
\end{equation}
where $\omega_k=2|\sin(\pi k/{N})|$ and $Q_k$ and $P_k$ are the
Fourier coefficients of $q_j$ and $p_j$. Then, assuming small
nonlinearity, we perform a near identity transformation to remove
nonresonant four-wave interactions (such procedure, is well documented
in the general case in \cite{falkovich1992kolmogorov} and in the
$\alpha$-FPUT case in
\cite{onorato2015}).  The following reduced Hamiltonian is obtained
(the new variable has been renamed $a_k$ and higher order terms have
been neglected):

\begin{equation}
\begin{split}
&\tilde H_{2}=N \sum_{k=0}^{N-1}\omega_k|a_k|^2 \\
  &\tilde H_4=\frac{N}{2}\beta \sum_{k_1 k_2 k_3 k_4}^{N-1} T_{k_1,k_2,k_3,k_4}a_{k_1}^*a_{k_2}^* a_{k_3} a_4 \delta_{1+2,3+4},
\end{split}
\end{equation}
{where all wave numbers $k_1, k_2, k_3$ and $k_4$ are
  summed from $0$ to $N-1$};
\begin{equation}
 T_{k_1,k_2,k_3,k_4}= \frac{3 }{4} e^{i \pi \Delta k/N}\prod_{j=1}^{4}
 \frac{2 \sin(\pi k_j/N)}{\sqrt{\omega(k_j)} }
\label{eq:coup_coef}
\end{equation}
with $\Delta k=k_1+k_2-k_3-k_4$ and $\delta_{i,j}$ is the generalized Kronecker Delta that accounts for a periodic Fourier space.
{
We then introduce scaled amplitudes $a_k'=a_k/\sqrt{H_2(t=0)/N}$  so that the equation of motion in the new variable read}:
\begin{equation}
i \frac{d a_{k_1}}{\partial t}=\omega_{k_1} a_{k_1}+
{\epsilon} \sum_{k_i}^{N-1}T_{k_1,k_2,k_3,k_4}a_{k_2}^*a_{k_3}a_{k_4}\delta_{1+2,3+4},
\label{zakh_eq}
\end{equation}
where primes have been omitted for brevity, the sum on $k_i$ implies a
sum on $k_2, k_3, k_4$ from 0 to $N-1$ and {
\begin{equation}
\epsilon = \beta H_2(t=0)/N,
\end{equation} 
that implies that our nonlinear parameter is proportional to the
linear energy density of the system at time $t=0$ and to the nonlinear
spring constant $\beta$.}
In terms of the angle-action variables
$
a_k=\sqrt{I_k}\phi_k\;\;\; {\rm with}\;\;\; \phi_k=\exp[-i \theta_k],
$
the equation for $\theta$ reads:
\begin{equation}
\begin{split}
&\frac{d \theta_{k_1}}{\partial t}=\omega_{k_1}
+ {\epsilon}\sum_{k_i}T_{k_1,k_2,k_3,k_4}\frac{\sqrt{I_{k_2}I_{k_3}I_{k_4}}}{\sqrt {I_{k_1}}}\times\\
&\ \ \ \ \ \ \ \ \Re[\phi_{k_1}^*\phi_{k_2}^*\phi_{k_3}\phi_{k_4}]\delta_{1+2,3+4},
\label{angle}
\end{split}
\end{equation}
where $\Re[...]$ implies the real part.  From this 
equation we obtain the frequency by
applying the averaging  operator $\langle...\rangle$ over random frequencies
and using the Wick's contraction rule,

\begin{equation}
\langle\phi_{k_1}^*\phi_{k_2}^*\phi_{k_3}\phi_{k_4}\rangle=\delta_{1,3}\delta_{2,4}+\delta_{1,4}\delta_{2,3},
\end{equation}
 we get the instantaneous frequency:
 \begin{equation}
\tilde \omega_{k_1}=\langle
\frac{d\theta_k}{dt}\rangle\simeq\omega_{k_1}+ {\epsilon}
2\sum_{k_2\neq k_1}T_{k_1,k_2,k_1,k_2}I_{k_2}
\label{eq:nonlindisprel}
\end{equation} 
i.e. the nonlinear dispersion relation given by the linear dispersion
relation plus amplitude corrections (recall that $I_k=|a_k|^2$), see
also \cite{gershgorin2005,gershgorin2007interactions}.  More
interestingly, one can estimate half-width $\Gamma_k$ of the frequency
by calculating the second centred moment of the equation (\ref{angle})
as:
\begin{equation}
\Gamma_{k}=\sqrt{\big\langle\left(\frac{d \theta_{k_1}}{\partial t}-\tilde \omega_{k}\right)^2\big\rangle}.
\label{eq:width}
\end{equation}
{
Using equations (\ref{angle}), (\ref{eq:nonlindisprel}), the  Wick's decomposition
and under the assumption of thermal equilibrium (equipartition of linear energy), 
we obtain:
\begin{equation}
\Gamma_k= \frac{3}{4}\epsilon  \omega_k=\frac{3}{4}\frac{1}{N} \beta
H_2(t=0) \omega_k
\end{equation}
} 
  
{
  Once the broadening of the frequency is estimated, the
   Chirikov overlap parameter can be defined as:
\begin{equation}
R_k=2 \frac{\Gamma_k}{\tilde\omega_{k+1}-\tilde\omega_{k}} \eqsim
\frac{3}{2} \frac{\omega_k}{\omega_{k+1}-\omega_k} \epsilon.
\end{equation}  
According to Chirikov, the stochastization takes place when
$R_k=1$. If we define $\epsilon_{\rm cr}$ as the value for which
$R_k=1$, then it is straightforward to observe that $\epsilon_{\rm
  cr}$ is $k$ dependent and $\epsilon_{\rm cr}$ becomes large for
small values of $k$. This implies that a transition region between two
regimes cannot be sharp. In the long wave limit the critical energy
takes the following form $H_{2\rm cr}(t=0)= {2 N}/({3\beta k})$.  Full
stochasticization of all wave numbers takes place for $\epsilon_{\rm
  cr}\simeq0.6$ (as we will see below, for this value of $\epsilon$ we
observe a new scaling of the equipartition time as a function of
time).  }

We now turn our attention to the estimation of the time scale needed
to reach equipartition.  The theoretical predictions that follows are
based on the assumption that an irreversible dynamics can be obtained
only if waves interact in a resonant manner, i.e. for some $n$ and $l$
the following system has solution for integer values of $k$:
\begin{equation}
\begin{split}
k_1+k_2+...+ k_l=k_{l+1}+k_{l+2}+...+k_n\\
\omega_{k_1}+\omega_{k_2}+...+ \omega_l=\omega_{l+1}+\omega_{l+2}+...+\omega_n.\\
\end{split}
\end{equation}
 Just like for a forced harmonic oscillator, non resonant interactions
 lead to periodic solutions, i.e. to recurrence. Based on the
 methodology developed in \cite{onorato2015}, we can state that for
 $N=32$ (the number of masses in original simulations of Fermi et al)
 there are four-wave resonant interactions; however, those resonances
 are isolated and can not lead to thermalization (see also
 \cite{henrici2008results,rink2006proof}). Following the results in
 \cite{onorato2015}, efficient resonant interactions for the
 $\beta$-FPUT take place for $l=3$ and $n=6$, i.e. six-wave resonant
 interactions is the lowest order resonant process for the discrete system.
  This implies that
 a new canonical transformation needs to be performed to
 remove non resonant four-wave interactions and obtain a deterministic
 six-wave interaction equation whose time scale is $1/\epsilon^2$, see
 \cite{laurie2012one} for details on the canonical transformation.

 An estimation of the time scale of such interactions can be obtained
 following the argument developed in \cite{onorato2015} based on the
 construction of an evolution equation for the wave action spectral
 density $ N_k=\langle |a_k|^2 \rangle$, $a_k$ being the new canonical
 variable \cite{onorato2015}.  Using the Wick's rule to close the
 hierarchy of equations, it turns out $ \partial N_k/\partial t\sim
 \epsilon^4$ The result is that the time of equipartition scales like
 $t_{eq}\sim 1 / \epsilon^4$ (this coincides precisely with the time
 scale, $1/\alpha^8$, given in \cite{onorato2015} for the
 $\alpha$-FPUT model).  In the continuum limit (thermodynamic limit)
 in which the number of particles $N$ and the length of the chain both
 tend to infinite, keeping constant the linear density of masses, then
 it can be shown that
 {
  the Fourier space becomes dense ($k \in
 \mathbb{R}$): four-wave exact resonances exists and
  the standard four-Wave Kinetic Equation can be recovered (see
  \cite{Spohn2006,lukkarinen2016kinetic,aoki2006energy,pereverzev2003fermi}). In this latter case the
  time scale for equipartition should be $1/\epsilon^2$.} 
  
{\it Numerical experiments-} We now consider numerical simulations of
the $\beta$-FPUT system in the original $q_j$ and $p_j$ variables to
verify our predictions.  We integrate the equations with $N=32$
particles using the sixth order symplectic integrator scheme described
in \cite{yoshida1990construction}.  We run the simulations for
different values of $\beta$, keeping always the same initial
conditions which is formulated in Fourier space in normal variables as
\begin{equation}
   a_k(t=0)=
    \begin{cases}
      e^{i \phi_k}/(N \sqrt{\omega_k}) & \text{if}\ k=\pm 1,\pm 2,\pm 3,\pm 4,\pm 5 \\
      0 & \text{otherwise},
    \end{cases}
 \end{equation}
related to the original variables by equation (\ref{NormalMode}).
$\phi_k$ are uniformly distributed phases. By changing $\beta$,
different values of the nonlinear parameter $\epsilon$ are
experienced.

\begin{figure}
\includegraphics[width=0.9\linewidth]{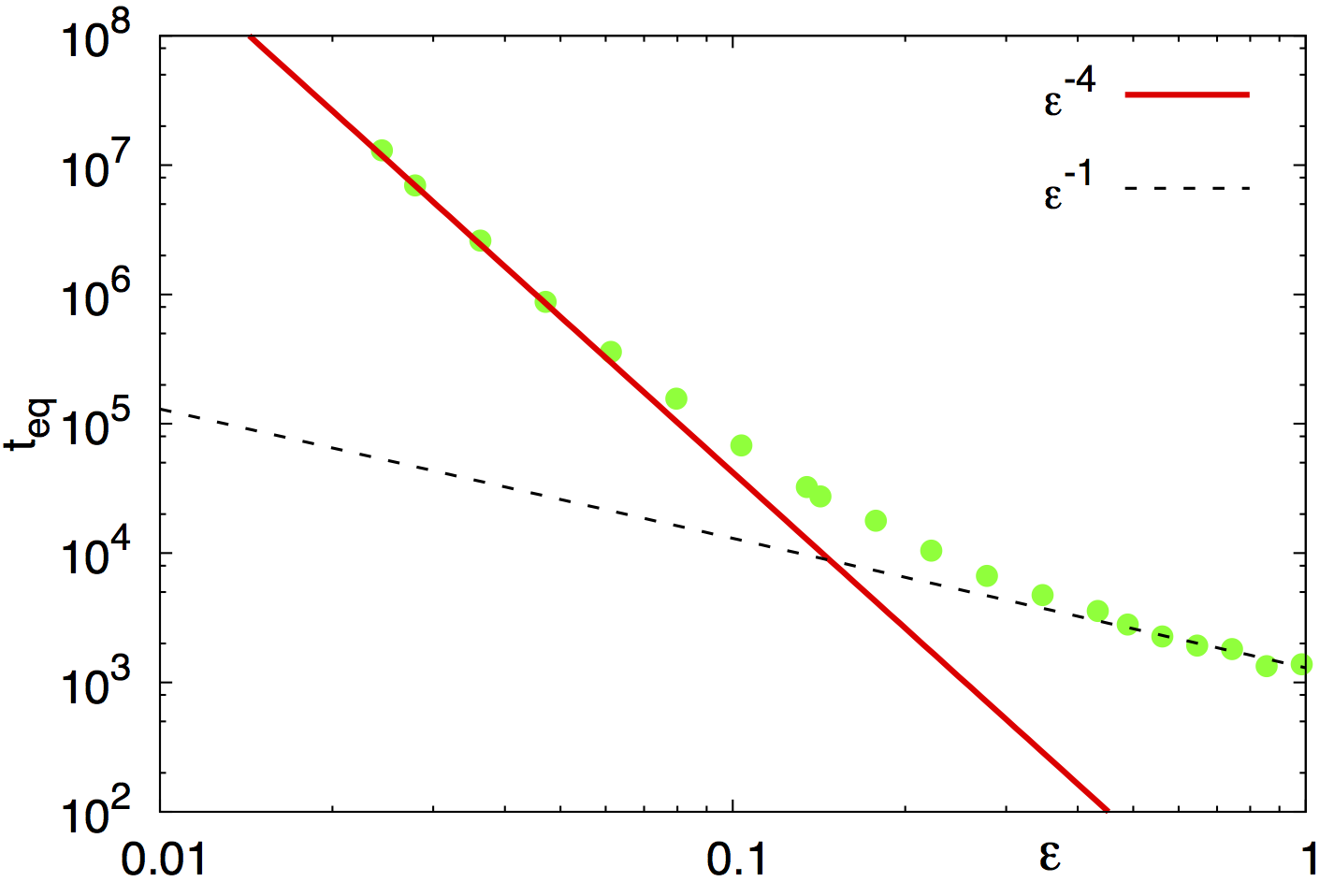}
\caption{Equilibrium time $t_{eq}$ as a function of $\epsilon$ in
  Log-Log coordinates.  Dots represent numerical experiments The
  straight line corresponds to power law of the type $1/\epsilon^{4}$,
  red line, and $1/\epsilon$, black dashed line.}
 \label{fig:scaling}
\end{figure}

We found out that a successful way of estimating the time of
equipartition is to run, for each initial condition, different
simulations characterized by a different set of random phases: our
typical ensemble is composed by 2000 realizations. In order to
establish the time of thermalization, we have considered the following
entropy \cite{livi1985equipartition}:
\begin{equation}
s(t)={-}\sum\limits_{k} f_k \log f_k\;\; {\rm with}\;\;\; f_k = \frac{N-1}{H_2}\omega_k \langle |a_k(t)|^2 \rangle,\;\;\; 
\end{equation}
and $\langle...\rangle$ defines the average over the realizations.
Note that the larger is the number of the members of the ensemble, the
lower is the stationary values reached by $s$. In our numerics we have
followed the procedure outlined in \cite{onorato2015}  to
identify the time of equipartition.  We present in Figure
\ref{fig:scaling} such time, $t_{eq}$, as a function of $\epsilon$ for
the simulations considered in a Log-Log plot.
\begin{figure}
\includegraphics[width=1\linewidth]{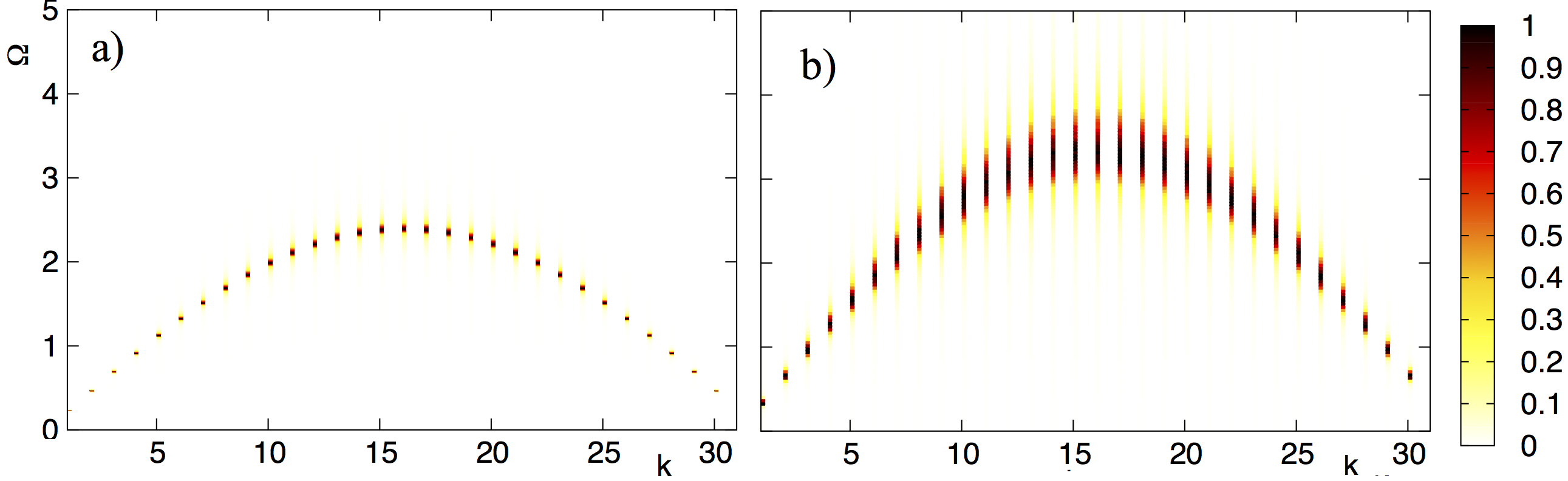}
\caption{$\langle |a(k,\Omega)|^2 \rangle$ for a) $\epsilon$=0.12 and
  b) $\epsilon$=1 obtained from numerical simulations.}
\label{fig:disp_rel}
\end{figure}
\begin{figure}
\includegraphics[width=1\linewidth]{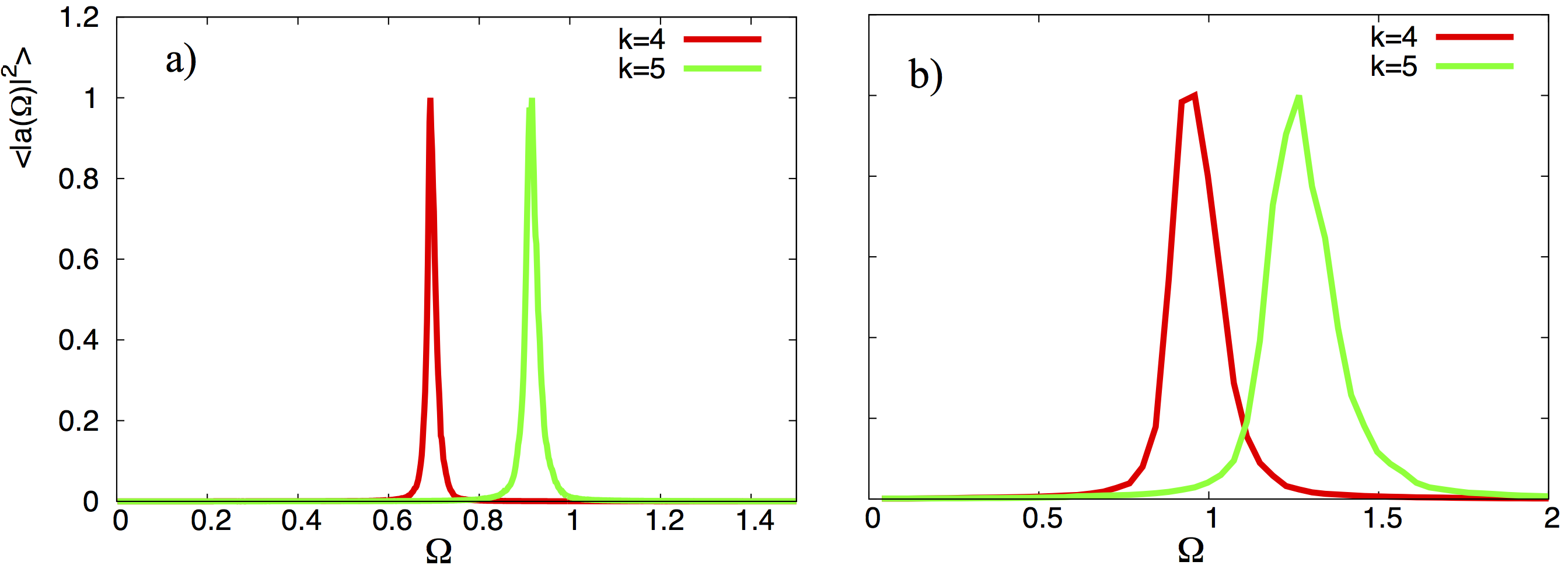}
\caption{$\langle |a(\Omega)|^2 \rangle$ for $k=4$ and $5$ for a)
  $\epsilon$=0.12 where there is no overlapping of the resonances of
  two nearby wave numbers and b) $\epsilon$=1 where there is 
a { noticeable} overlapping of the resonances of the nearby wave numbers}
\label{fig:freq_overlap}
\end{figure}
The figure also shows two straight lines (power laws) with slope -4
and -1. The steepest one (in red color) is consistent
with the six-wave interaction theory, while the blue one corresponds
to the time scale associated with the nonlinearity in the dynamical equation.
 A clear transition between the two scalings is observed.
Similar transition has also been observed in \cite{Danieli2017} where
the $\alpha$-FPUT has been integrated.

In order to understand such behaviour, we build the dispersion
relation curve from numerical data and measure the shift and the width
of the frequencies as a function of the parameter $\epsilon$. After
reaching the thermalized state, the procedure adopted consists in
constructing the variable $a_k(t)$ from eq. (\ref{NormalMode}) and let
the simulation run on a time window over which, for each mode, a
Fourier transform (from variable $t$ to $\Omega$) is
taken. This is done for all the members of the ensemble.  Then $\langle
|a(k,\Omega)|^2 \rangle$ is  normalized by its maximum for each
value of $k$ and then plotted as a function of $k$ and $\Omega$.
\begin{figure}
\includegraphics[width=0.8\linewidth]{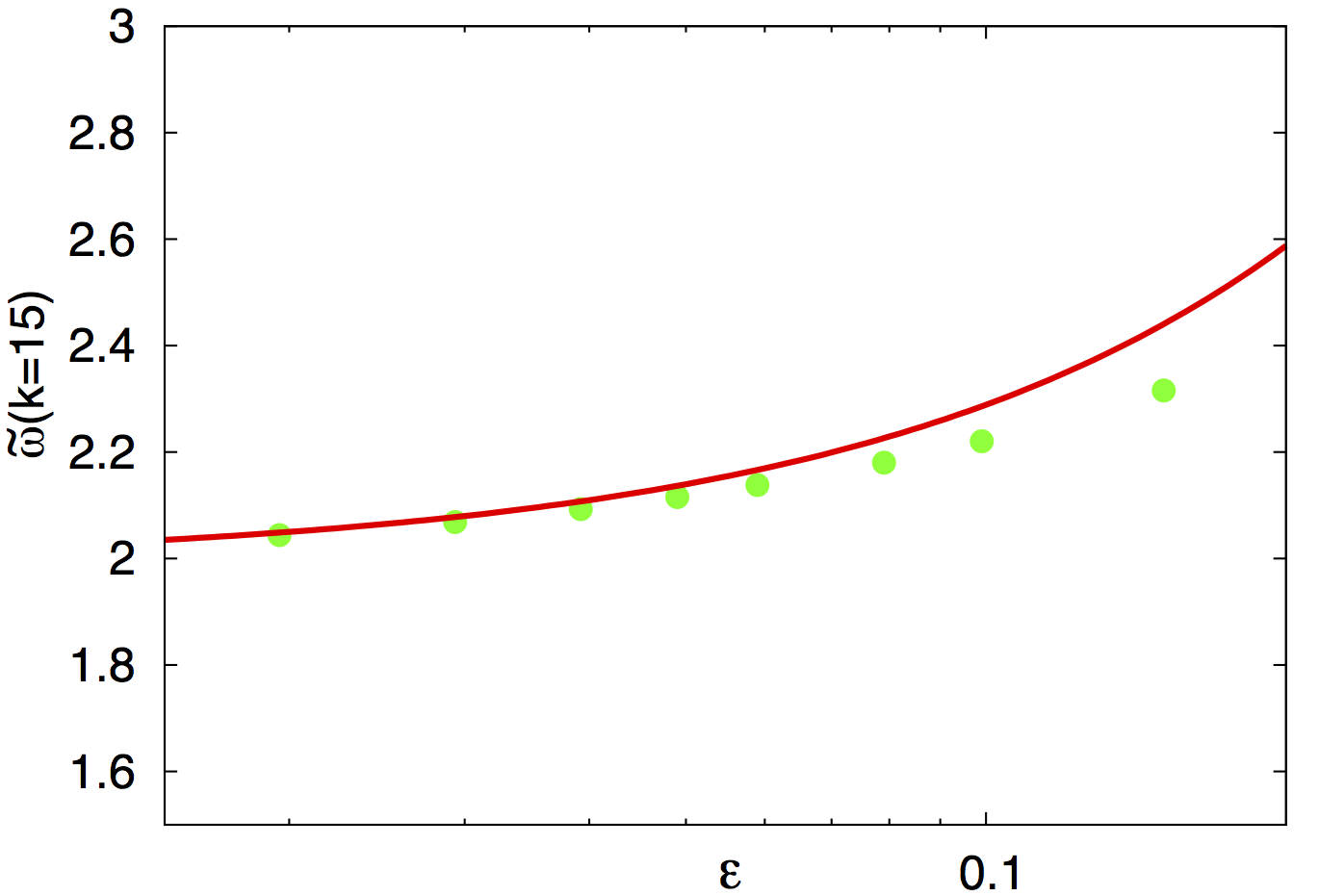}
\caption{Frequency shift as a function of the nonlinear parameter
  $\epsilon$ for $k$=15. The solid line corresponds to equation
  (\ref{eq:nonlindisprel}), dots correspond to the position of the
  peak of the distribution $\langle |a(\Omega)|^2 \rangle$ for $k=15$
  computed numerically.}
\label{fig:shift}
\end{figure}
\begin{figure}
\includegraphics[width=0.8\linewidth]{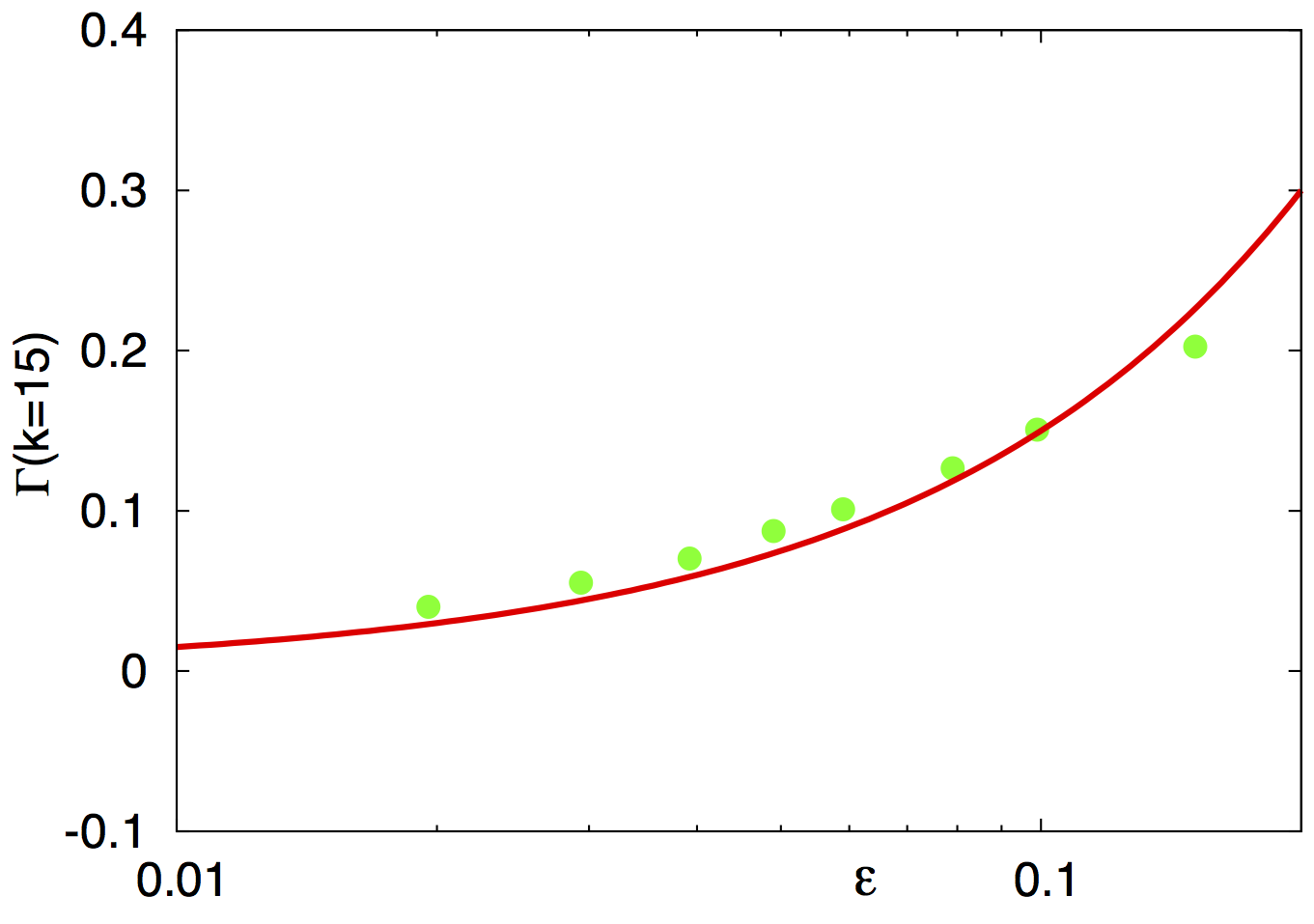}
\caption{Width of the distribution of the frequencies shift as a
  function of the nonlinear parameter $\epsilon$ for $k$=15. The solid
  line corresponds to equation (\ref{eq:width}), dots correspond to
  the standard deviation of the distribution $\langle |a(\Omega)|^2
  \rangle$ for $k=15$ computed numerically.}
\label{fig:width}
\end{figure}
Would the system be linear, only discrete Kronecker Deltas would
appear, placed exactly on the linear dispersion relation curve,
i.e. $\langle |a(k,\Omega)|^2 \rangle=\delta_{\Omega,\omega_k}$.  In
Figure \ref{fig:disp_rel} we show two examples of the ($\Omega-k$)
plot: the first is calculated on the transition region,
$\epsilon=0.12$, and the other one in stronger nonlinear regime,
$\epsilon=1$. The plots appear to be very different:
first we  notice that for the stronger
nonlinear case the dispersion curved is shifted towards higher
frequencies.  The shift is  less pronounced for the smaller
nonlinearity case, $\epsilon=0.12$ (in the linear case, the curve
touches $\Omega=2$). 

The other important aspect is that a { noticeable} frequency broadening is
observed for $\epsilon=1$; that implies that for a single wavenumber,
there is a distribution of frequencies characterized by some
width. Due to such width, for two adjacent discrete wavenumbers, the
frequencies overlaps (Chirikov criterium). In order to have a clearer
picture of such overlap, we show a slice of Figure \ref{fig:disp_rel}
taken at $k=4$ and $k=5$ for both cases, see Figure
\ref{fig:freq_overlap}. The distribution of the frequencies are 
separated for the weakly nonlinear case and visibly overlap for the
stronger nonlinear case. Note that also for the weakly nonlinear case,
for larger wave numbers an overlap starts to appear (not shown in the
figure). This is the reason why the prediction made on exact six-wave
resonant interaction starts failing and an other scaling is observed
(see Figure \ref{fig:disp_rel}).
  
We compare the shifts and the broadening of the frequencies of our
theoretical predictions with the one obtained from numerical
simulations, see Figures \ref{fig:shift} and \ref{fig:width}. Results
are overall in agreement in the very weak nonlinear regime: the
predictions are obtained by assuming the random phase approximation
which does not hold as soon as the nonlinearity starts creating
correlation between wave numbers. Such departure of the theory is
consistent also with the one observed in Figure \ref{fig:scaling}.

{\it Conclusions-} In this Letter we have considered the original
$\beta$-FPUT model and found that the system reaches a thermalized state,
even for very small nonlinearity.  In this regime and for small number of
modes, three time scales may be identified: the linear time scale
$1/\omega_k$, the nonlinear time scale of four wave interactions, and
the time scale of irreversible six wave interactions,
$1/\epsilon^4$. In order to observe equipartition one needs to wait up
to the $1/\epsilon^4$ time scale. If one is observing the system on a
shorter time scale, using the original variables, then only reversible
dynamics is seen, which might be an explanation for the celebrated
FPUT recurrence.  Such reversible dynamics can be possibly captured
directly as is done in \cite{guasoni2017incoherent}, where a
nonequilibrium spatiotemporal kinetic formulation that accounts for
the existence of phase correlations among incoherent waves is
developed.

For $\epsilon\gtrsim0.1$ a different scaling, $t_{eq}\sim \epsilon^{-1}$, starts
which is consistent with the time scale of the nonlinearity
of the dynamical equation. The transition region has been
investigated by measuring the broadening of the frequencies: we
observe that the phenomenon of frequency overlap suggested by Chirikov
starts in the transition region; the breakdown of the prediction of
the discrete weak wave turbulence theory is then observed for such
values of nonlinearities. {Chirikov criterion approximately
  separates regimes of slow equipartition due to six-wave
  resonant interactions to the other, more effective mechanism for
  reaching thermal equilibrium.  } The mechanism that leads to
equipartition for weakly nonlinear initial conditions seems to be
universal; indeed, the $\alpha$-FPUT behaves exactly in the same way
and we expect that the same mechanism be responsible for explaining
the equipartition in systems where metastable states has been observed
as in the Nonlinear Klein-Gordon equation \cite{fucito1982approach}.

{\bf Acknowledgments} M.O. has been funded by Progetto di Ricerca
d'Ateneo CSTO160004 . The authors are grateful to D. Proment and
Dr. B. Giulinico for discussions.


\end{document}